\documentstyle[12pt]{article}
\newcommand{\newc}{\newcommand}
\newc{\zeus}{{\sc Zeus}}
\newc{\gev}{\,GeV}
\newc{\rp}{$R_p$}
\newc{\rpv}{$\not\!\!R_p$}
\newc{\rpvm}{{\not\!\! R_p}}
\newc{\rpvs}{{\not R_p}}
\newc{\ra}{\rightarrow}
\newc{\lsim}{\buildrel{<}\over{\sim}}
\newc{\gsim}{\buildrel{>}\over{\sim}}
\newc{\esim}{\buildrel{\sim}\over{-}}
\newc{\lam}{\lambda}
\newc{\lsp}{{\tilde\Lambda}}
\newc{\chio}{{\tilde\chi_0}}
\newc{\oc}{{\cal {O}}}
\newc{\msq}{m_{\tilde q}}
\newc{\pho}{{\tilde\gamma}}
\newc{\dbar}{{\bar d}}
\newc{\ubar}{{\bar u}}
\newc{\half}{\frac{1}{2}}
\newc{\third}{\frac{1}{3}}
\newc{\quarter}{\frac{1}{4}}
\newc{\beq}{\begin{equation}}
\newc{\eeq}{\end{equation}}
\newc{\barr}{\begin{eqnarray}}
\newc{\earr}{\end{eqnarray}}
\newc{\ptmis}{\not\!\!{p_T}}
\newc{\mc}{\multicolumn}

\newc{\mpl}{M_{Pl}}
\newc{\mw}{M_W}
\begin{document}

\title{{\bf SIGNALS FOR SUPERSYMMETRY AT HERA}}
\author{Herbi DREINER\thanks{This talk is a summary of work done in
collaboration with Jon Butterworth and Peter Morawitz.}\\
{\em Theoretische Physik, ETH-Z\"urich, 8093 Z\"urich, Switzerland}}

\maketitle
\setlength{\baselineskip}{2.6ex}

\begin{center}
\parbox{13.0cm}
{\begin{center} ABSTRACT \end{center}
{\small \hspace*{0.3cm}
We summarize two sets of signals at HERA in supersymmetric  models with
explicitly broken R-parity. (1) The resonant production of single squarks
through an operator $L_1Q_i{ \bar D}_j$. (2) The MSSM production of a squark
and a slepton followed by
R-parity violating decays via the operators $L_iQ_j{\bar D}_k$. The first
case allows for a very large mass reach: $m_{\tilde q}\leq 275\gev$ for
moderate couplings. For masses $\sim 100\gev$ couplings down
to $\lam\geq10^{-3}$ can be probed. In the second case the production mechanism
is independent of the R-parity violating Yukawa couplings.
The full generation structure and
very low couplings, $10^{-6}$(!!), can be probed. In the most optimistic case
the mass reach is $170,195,205\gev$ for the operators $L_{\tau,\mu,e}Q{\bar
D}$ respectively. We conclude that HERA offers a {\it very promising}
discovery  potential for \rpv\  SUSY.}}
\end{center}

\section {Introduction}
When extending the Standard Model to include supersymmetry \cite{susyreview}
the superpotential contains terms  which lead to unsuppressed proton decay. The
most elegant and economic solution to this problem is to impose a discrete
symmetry which disallows the dangerous $\Delta B \not=0$ terms. Several
discrete symmetries have been proposed, with R-parity ($R_p$)
and Baryon Parity ($B_p$) being the most common. Imposing $R_p$ results in
the minimal supersymmetric standard
model (MSSM); in the case of $B_p$ the superpotential contains in addition
the following lepton number violating terms\cite{rp1}\footnote{The notation is
as in \cite{butter}.}
\beq
\lam_{ijk}' L_iQ_j{\bar D}_k,\quad \lam_{ijk} L_iL_j {\bar E}_k.
\label{eq:operators}
\eeq
These terms can have significant cosmological consequences \cite{rpcosmo},
however here we only focus on collider phenomenology. For $e^+e^-$-colliders
the effects have been studied in \cite{rpee} and for hadron colliders
in \cite{rphadron,rphadronsum}.
In this talk we consider the phenomenology of the first set of terms
at HERA: $\lam'L_iQ_j{\bar D}_k$. We consider two sets of signals. First
the resonant production of squarks followed by the cascade decay to the
lightest supersymmetric particle (LSP)
\beq
e^-+ u_i  \ra ({\tilde d}_{jR})^* \ra d_{jR}+\pho, \quad
e^-+ {\bar d}_j  \ra ({\tilde{\bar u}}_{iL})^* \ra {\bar u}_i +\pho ,
\label{eq:two}
\eeq
This production mechanism was first studied by Hewett \cite{joanne}. Second,
we consider the MSSM production\cite{heraws},
\beq
e^-+q\ra {\tilde e}^- + {\tilde q}, \quad
e^-+q \ra {\tilde \nu} + {\tilde q}'. \label{eq:ccprod}
\eeq
followed by \rpv-decays of the sfermions
${\tilde f}\ra f + {\tilde \chi}^0_1$ $ \ra f + \{(e^\pm_i+u_j+{\bar d}_k)
{\mbox{ or  }}$ $ (\nu_i+d_j+{\bar d}_k) \}.$
where ${\tilde \chi}^0_1$ denotes the neutralino LSP.
The first mechanism leads to a large mass reach the second to a large
reach in coupling constant. In both cases the main signal is a high-$p_T$
positively charged lepton.

In our analysis we shall make the following simplifying assumptions
{\bf (A1)} The LSP is a neutralino. In some regions of the MSSM
parameters the LSP is a chargino. We do not study these supersymmetry models.
{\bf (A2)} Of the 27 operators $L_iQ_j{\bar D}_k$ we shall assume in
turn that only one lepton number is violated
\cite{rphadron}. We thus have 3 separate scenarios to
consider. {\bf (A3)} The cascade decay of the sfermions to the LSP is
100\% of the branching fraction.

\section{Previous Bounds}

The existing bounds on the operators $\lam_{ijk}' L_iQ_j{\bar D}_k$
are of two kinds: (a) indirect bounds from low-energy processes involving
virtual supersymmetric particles; the most stringent being \cite{vernon}
$\lam_{L_eQ_i{\bar D}_j}' \leq
0.06\, ( {{\tilde m}}/{200\gev}),$ {\it i}=1;
$ 0.52\,  ({{\tilde m}}/{200\gev}),$  {\it i}=2,3;
$\lam_{L_\mu Q_i{\bar D}_j}' \leq
0.18\,  ({{\tilde m}}/{200\gev}),$ {\it i}=1;
$ 0.44\,  ({{\tilde m}}/{200\gev}),$ {\it i}=2,3.
(b) Collider bounds; the most detailed  have been determined by D.P. Roy
\cite{dp} using the old dilepton search data of CDF.
The bound on the squark masses thus obtained is
${\tilde m}_q \geq 100\gev,$ for
$L_eQ_i{\bar D}_j ,\,L_\mu Q_i{\bar D}_j.  $
These bounds are currently being updated with the new data.
Note there are {\bf no}  bounds on the tau-operators $L_\tau Q_i {\bar D}_j.$
In addition there are model independent
bounds from LEP, which hold for all operators
\cite{lepbound}
${\tilde m}_{e}\geq45\gev,$ ${\tilde m}_q \geq 45\gev.$
Thus our best bounds are
\beq
{\tilde m}_e+ {\tilde m}_q \geq
145\gev, \quad L_{e,\mu}Q{\bar D},
\qquad {\tilde m}_e+ {\tilde m}_q\geq90\gev,\quad \,\,\,\, L_{\tau}Q{\bar D}.
\eeq

\section{Resonant Squark Production}
To determine the production rate we fold the parton-level cross section
with the quark structure functions and then add the decay of the real
LSP, limiting ourselves to the positron decay mode. For $\lam'=0.03$ the
event rate is typically $10^3/200\,pb^{-1}$ for a $120\gev$ squark and
drops to 1 for a $260\gev$ squark. These rates scale with $lam'^2$.

\begin{table}
\begin{center}
\begin{tabular}{||c|c||c|c|c|c||} \hline
{ Variable} & { Cut }   & \multicolumn{4}{||c||}{Events Remaining} \\
\cline{3-6}
{ used }      & {     }   & {\rpv}  & {$b\bar{b}$} & {NC DIS} & {$W^+$} \\
\hline
{None}          & $-$  & $152$ & $5.1\times 10^5$ & $2.6\times 10^5$ & $16$\\
$p_{Te} (GeV/c)$     & $>2$   & $63$  & $12700$      & $1550$   & $13$\\
$E_T (GeV/c^2)$      & $>120$ & $56$  & $0$          & $26$     & $0.7$\\
$E_{TOTAL} (GeV/c^2)$ & $>450$ & $56$  & $0$          & $0$     & $0.2$\\
\hline
\end{tabular}
\caption{Cuts (2) and Efficiencies, $M_{SUSY}=240\gev$ }
\end{center}
\end{table}

The events typically have a high $p_T$ positron, and also large values of
total transverse energy ($E_T$), since we expect the photino and the squark
to be relatively massive. We also make use of the total energy in the detector
which is a good variable for the very massive squarks, and a slight missing
$E_T$ cut to reduce theW-production background.

We have considered two types of physics background in detail. The first is
b-quark production $via$ photon-gluon fusion. These are generally low-$E_T$
events, but they have
relatively high cross sections at HERA and thus their high $E_T$ tail will
present a significant background. The HERWIG generator \cite{herwig} was used
to generate a
sample of $b$-quark production events.
The cross section for this process
was estimated to be $2430~pb$ with the KMRS B0 set of structure functions
\cite{KMRS}.
The second background considered is $W^+$ production
and the $W^+$ can decay to give a positron. A sample of these events was
generated with the generator used in \cite{zep1}. The cross section for this
process multiplied by the branching fraction for the ($W^+ \ra e^+  \nu_e$)
was estimated to be $78~fb$.

\lq Fake\rq\ backgrounds are those caused not by genuine positron production,
as in the processes above, but by effects inside the \zeus\ detector faking a
high $p_T$ positron signal. The dominat effect is the
misidentification of the sign of the curvature (and thus charge) of high
$p_T$ electrons in the tracking detector.
In order to investigate this a sample of high $Q^2$ deep inelastic
scattering neutral current events were generated, using the program LEPTO
\cite{LEPTO}. These events provide a source of high $p_T$ electrons.  The
kinematic region over which these events were generated was
$300\,GeV^2 < Q^2 < 9.84\cdot 10^{4}\, GeV^2 ,\quad
10^{-4} < x < 0.9.$
The cross section within this region is $1350~pb$ for the KMRS B0 set of
structure functions.

In order to get an idea of our analysis I have included Table 1 for
the extreme case of a $240\gev$ squark. Note that for the $b{\bar b}$
and NC-DIS a modest parton level $E_T$ cut was imposed before the
full simulation in order to reduce the computational effort.

Applying the discovery limit criteria: {\bf (1.)}
Five or more \rpv\ events surviving our cuts per nominal HERA year ($
200pb^{-1}$),
{\bf (2.)} A signal/$\sqrt{\rm background}$ ratio of at least three,
to these plots we calculate that \rpv\ should be observable in the \zeus\
detector
up to a squark mass of
\beq
M_{\tilde{q}} \lsim 270 \,GeV,
\eeq
for $\lam'_{11j} = 0.081$. For $M_{\tilde{q}} = 100\,GeV$, \rpv\ should be
observable for
\beq
\lam'_{11j} \gsim 0.0053
\eeq
Carrying out a similar procedure for the $\lam'_{12i}$ cases we
find correspondingly smaller discorvery reaches since the structure functions
are smaller for the higher generations. With this years expected data of
$5\,pb^{-1}$ one should be able to probe squark masses upto about $180\gev$.

\section{MSSM Production}
The production mechanisms have been discussed elsewhere in detail
\cite{heraws}. The decay of the sfermions is completely analogous
to that of the squarks in the previous section.
The main difference in this analysis is that we have considered a
general neutralino LSP, not just a photino. This leads to
two important effects. {\bf (1.)} The charged lepton branching fraction
of the LSP strongly depends on the composition of the neutralino.
It can vary from $70\%$ for a pure photino down to $15\%$ in the
extreme. {\bf (2.)} For a Higgsino/Zino dominated LSP the mass
can become so small that the LSP no longer decays within the detector.
Then the MSSM analysis applies.
In order to reduce the large SUSY parameter
space, we have made a number of simplifying assumptions
{\bf  {(1)}}
We have assumed sleptons and the five
squarks to be degenerate in mass: M$_{SUSY}=m_{\tilde l}=m_{\tilde q}$.
{\bf {(2)}}
We have only considered two sets of gaugino mixing parameters: ($M'=40\gev$,
 $\tan(\beta)=1$ and $\mu=-200\gev$) where $BF(LSP\ra l^\pm+X)\gsim65\%$
and ($M'=55\gev$, $\tan(\beta)$=4 and $\mu=+200\gev$)
where $BF(LSP\ra l^\pm+X)\lsim20\%$.

{\small{
\begin{table}
\begin{center}
\begin{tabular}{c|rrrrr}
 cut  &  $b {\bar b}$ & $c {\bar c}$ &  W
 &  NC &
CC      \\
\hline
\hline
none &    3395 & 4105 & 400 & 7149 & 3541 \\
$e^+$ or $\mu$ found &   148 &  95 &  183 &
84 & 19 \\
$30\gev < (E-p_z) < 60\gev$ &    134 & 80 &  51 & 74
& 5 \\
circularity  $>$ 0.1 &     46 & 25 & 12 & 7 &
0 \\
$E_T > 90\gev$ &         2 & 0 & 0 & 0 & 0 \\
\hline
\hline
Exp. no. of evts &   0.53 $\pm$ 0.31&  0 & 0 & 0 & 0 \\
\end{tabular}
\end{center}

\begin{center}
\begin{tabular}{c|rrrrr}
 cut  &  $L_\tau Q {{\bar D}}_{45}$ & $L_\tau Q {{\bar D}}_{65}$ & $L_\tau Q
{{\bar D}}_{80}$ & $L_\tau Q
{{\bar D}}_{85}$ & $L_\tau Q {{\bar D}}_{90}$ \\
\hline
\hline
none &  2000 & 1000 & 1000 & 400 & 200 \\
$\mu$ found & 185 & 74 & 92 & 33 & 14 \\
$30\gev < (E-p_z) < 60\gev$ & 166 & 53 & 71 & 27 & 11 \\
circularity  $>$ 0.1 & 105 & 40 & 58 & 21 & 8 \\
$E_T > 90\gev$ &        17 &  26 & 44 & 17 & 8 \\
\hline
\hline
Exp. no. of evts & 17.2 $\pm$ 4.2 & 13.4 $\pm$  2.6 & 6.6 $\pm$  0.9 &
3.9 $\pm$  0.9 & 2.2 $\pm$  0.8 \\

\end{tabular}
\end{center}
\caption{\label{t:lqd311} The cut-sequence applied to the MC samples.
 Here ``$L_\tau Q {{\bar D}}_{45}$'' refers to the  SUSY  sample with
M$_{SUSY}=45\gev$, and LSP decays via the $L_\tau Q {{\bar D}}$ operator.}
\end{table}}}

{\it Signatures and Background}:
{}From  the processes (\ref{eq:ccprod}) we obtain 0,1, or 2 charged leptons
$e,\mu$, or $\tau$. The $\tau$ can subsequently decay to an $e$ or a $\mu$.
Our overall signals for LSP decays via the $L_\tau Q{\bar D}$ are thus a
positron or a muon and  jet activity in the detector.
For the $L_e Q{\bar D}$ and the $L_\mu Q{\bar D}$ operator, our
signals are one or more positrons {\it or} one or more muons respectively
and   jet activity in the detector.
The background is analogous to the previous section with the following
distinctions: (a) we now include $c{\bar c}$-production. This has a small
$E_T$ distribution but a very large cross section; the effect is
comparable to $b{\bar b}$-production. (b) for the $\mu$-signals we
consider the $\mu$-decays of the heavy flavours and of the $W^+$.
The resulting events have a different topology than those considered
in section 3; for example $p_T$ is no longer a good discriminator
between signal and background. We thus perform the following cuts:

{\bf (1)} In order to keep the monte carlo study manageable we perform a
generator-level cut of $Q^2> (22.36)^2{\gev}^2$ on
the background samples. This corresponds to an $E_T>45\gev$ cut which
is significantly lower than our final $E_T$ cut. {\bf (2)} Require an
$e^+$ or a  $\mu^\pm$ found. {\bf (3)}
30\ GeV$< (E-p_z) < 60$\ GeV: $E$ is the sum of the total energies
 of the final state particles measured in the calorimeter,
 and $p_z$ is the  sum of the z component of
 momentum \footnote{The z-axis is defined to point along the proton
direction.}  of   all particles measured in the calorimeter.
In our study we have used the $(E-p_z)$ cut to mainly reject charged current
(CC) events and W events. In both types of background a considerable
fraction of the energy is carried off by neutrinos, and hence will not be
detected. {\bf (4)} circularity $>$ 0.1: One can define a 2-dimensional-like
sphericity, circularity \cite{circularity}, which is defined in the
range $0<$circularity$<1$. The circularity   reflects the isotropy of the
event. A  circularity of unity corresponds to a completely isotropic  event.
Due to the high jet activity the  SUSY events  are
very  isotropic, and thus have a high value of circularity.
{\bf (5)} $E_T >$ 90\ GeV: $E_T$   is the total transverse energy measured
in the calorimeter. We also include the measurement of the muon track
in the  $E_T$  if a muon was found.

We determine the discovery potential using the same criteria as in the
previous section. We find for the high and low $BF$ scenario respectively:
\newline $m({\tilde e}, {\tilde \nu})+m({\tilde q})\leq$
\barr
\left\{
\begin{array}{ccc}
170\gev,& 195\gev,&205\gev \\
140\gev,& 175\gev,& 185\gev
\end{array}\right\}&&
for\, L_{\tau} Q {{\bar D}}, L_{\mu} Q {{\bar D}}, L_{e} Q {{\bar D}}
\earr
after two years of   HERA running at nominal luminosity. HERA is the first
collider experiment which can directly probe the $L_\tau Q{\bar D}$.

We conclude that HERA offers a very promising  discovery potential for
direct  searches
for  \rpv\    SUSY. For the resonant production new regions will be probed
with this year's data. For the MSSM production about $5\,pb^{-1}$ are required
to probe the $L_{\tau}QD$ operator, and at least $20\,pd^{-1}$ to probe
$L_{e,\mu}QD$.

\bibliographystyle{unsrt}

 \end{document}